\documentclass{aa}

\begin{document}
\input{psfig}
%   \thesaurus{06     % A\&A Section 6: Form. struct. and evolut. of stars
%              (03.11.1;  % Cosmogony,
%               16.06.1;  % Planets and satellites: general,
%               19.06.1;  % Solar system: general,
%               19.37.1;  % Stars: formation of,
%               19.53.1;  % Stars: oscillations of,
%               19.63.1)} % Stars: structure of.
%
   \title{Infall variability in the Classical T Tauri system VZ Cha}

%   \subtitle{or at least two of them}

   \author{Kester Smith\inst{1,2}
          \and
          Geraint F. Lewis\inst{3}
          \and
	  Ian A. Bonnell\inst{4}
          \and
          James P. Emerson\inst{5}
          }

   \offprints{K. Smith}

   \institute{Institut f\"{u}r Astronomie, ETH-Zentrum,
              CH-8092 Z\"{u}rich, Switzerland. 
           \and
             Paul Scherrer Institut, W\"{u}renlingen und Villigen,  
              CH-5232 Villigen PSI, Switzerland. \\
              kester@astro.phys.ethz.ch \\
           \and
              Anglo-Australian Observatory, P.O. Box 296, Epping, 
              NSW 1710, Australia \\
              gfl@aaoepp.aao.GOV.AU \\
           \and
              University of St Andrews, St Andrews, Scotland. \\
              iab1@st-andrews.ac.uk \\
           \and
              Department of Physics, Queen Mary, University of London,
              Mile End Road, London E1 4NS, UK. \\      
	      j.p.emerson@qmw.ac.uk \\
             }

   \date{Received June 2001, ; accepted , }

\abstract{We present time series spectroscopy of the Classical T Tauri
star VZ Cha. We follow spectral variations at intermediate resolution
over five successive nights, or approximately two rotation periods. We
see profile features which persist on timescales longer than the
expected infall time from the inner disc, and we see expected evidence
of rotational variations in the lines, but we also note that rotation
alone cannot produce all the observed variability and some other
mechanism must be invoked. The behaviour of H$\alpha$ is observed to
be markedly different from that of the other lines. In
particular, the evidence of rotational effects is lacking at
H$\alpha$, and the activity in the red and blue wings of the line is
not significantly 
correlated, in contrast to the other Balmer lines.  \keywords{Stars:
T Tauri -- stars: accretion -- stars: spectroscopy }}

\maketitle

%
%________________________________________________________________

\section{Introduction}

Classical T Tauri stars (CTTS) accrete material through circumstellar
discs at a rate of $10^{-8}$ to $10^{-7}$ M$_{\odot}$yr$^{-1}$
(Gullbring et al 1998). This accretion leads to strong infrared and
ultraviolet excess continuum emission, and the formation of emission
lines. Absorption features relative to the mean profile are often seen
in the red wings of emission lines (Edwards et al 1994) and indicate
free-falling material in the line of sight (e.g. Bonnell et al,
1998). Coupled with various other factors such as generally rapid
rotation rates, apparent star-disc rotation locking and the occasional
presence of photospheric spots, these suggest that the inner part of
the circumstellar disc is truncated by a strong stellar magnetic field
and the accreting material is channelled down the field lines onto the
star (K\"onigl, 1991). This picture is analogous to models developed
for magnetic neutron star systems (e.g. Ghosh \& Lamb 1978 and
following papers), and became more firmly established for CTTS with
the detection of kilogauss-strength magnetic fields in small samples
of T Tauri systems (Basri et al, 1992, Guenther et al 1999,
Johns-Krull et al 1999).  The geometry and detailed nature of the
magnetically funnelled accretion process remain poorly understood, and
are of importance if we are to understand both accretion and related
phenomena such as jets and outflows or the regulation of the star's
angular momentum during pre-main sequence evolution. A crucial initial
question is whether the field is a large-scale ordered dipole,
presumably tilted with respect to the stellar rotation axis, or
whether it has more complex structure, perhaps similar to scaled-up
magnetic loops in the Solar corona. In the former case we expect to
see either one or two distinct accretion streams pass through the line
of sight in each rotation period. In the latter case, there could in
principle be any number of loops. Solar active regions can persist for
weeks and by analogy we would expect T Tauri magnetic structures to survive
at least for timescales comparable to the rotation period.

Modelling the optical spectroscopic appearance of these regions, which
will be ultimately necessary for comparison with spectral variations,
is problematic as full radiative transfer models are needed. Line
profile modelling has been undertaken in an attempt to reproduce the
observed hydrogen line profiles (Hartmann et al 1994, Muzerolle et al
1998). Line profiles were modelled as being due to the velocity
distribution of the infalling material in the magnetosphere. The
profiles produced were in good agreement with observed profiles,
although no wind component was considered and Stark broadening was
needed to explain the very broad profile of H$\alpha$. Temperatures in
the magnetosphere would need to be of the order of 7000 - 10,000 K,
and the heating mechanism required is not specified, although magnetic
effects (i.e. reconnection) would be a reasonable assumption.

The line profiles, and in particular the redshifted absorption
components, are typically found to be strongly variable, and this is
usually explained as being caused by accretion streams crossing the
observer's line of sight as the system rotates. This implies an
azimuthally-asymmetric accretion geometry. The simplest such geometry would be
a dipole inclined to the stellar rotation axis, a situation
which might be expected in any case by analogy with the Sun.  Together with
knowledge of the rotation period provided by previous photometric
work, these variations could in principle be used to reveal the
detailed structure of the magnetically dominated inner regions. It
seems that non-axisymmetric models are applicable in at least some
cases. For example, Kenyon et al (1994) observed photometric
variability of the extremely active CTTS DR Tau which was modelled as
being due to a hot spot on the surface. This spot was identified as
the accretion shock at the base of an accretion stream, indicating
either a tilted dipole or a non-global field
geometry. Hessman and Guenther (1997) observed quasi-periodic
variations in the fluxes of the major emission lines of DR Tau and DG
Tau, attributed also to rotational modulation of the emitting
accretion shock. Smith et al (1998) also observed DR Tau, and reported
varying redshifted absorption, consistent with one of several
previously detected photometric periods and therefore suggestive of a
rotating accretion stream being carried through the observer's line of
sight.  Analysis of equivalent width variations from the same data set
revealed time lags of several hours between the four lowest Balmer
lines and Ca II H and K, with the higher Balmer lines varying before
the lower ones (Smith et al, 1999). In the rotating-streams scenario,
this would indicate that the variations of the different Balmer lines
are linked but spatially displaced in some way.  The most likely
explanation is that the proportion of flux in the Balmer lines varies
along the accretion stream, with the higher lines having a larger
proportion from close to the hot accretion shock on the stellar surface. 
There is some evidence for
this from the line flux modelling of Muzerolle et al (1998). The
interpretation proposed was that the time lags were the signature of a
stream moving over the limb, the hot dense regions near the star being
seen first with decreasing occultation by overlying stream material, then
disappearing over the limb.

Analysis of line profiles at high resolution for SU Aur was made by
Johns and Basri (1995). They reported evidence of both infalling and
outflowing components, which seemed to vary periodically with a
180$^{\circ}$ phase difference. The interpretation was an
``egg-beater'' model in which a rotating inclined stellar dipole field
disrupts a disk and is responsible for powering a disk wind. This
model resembles models proposed by Shu et al (1994) and also Paatz and
Camenzind (1996) in which a dipole stellar field powers an outflow
from the inner disc.  A major outstanding question is how these disc
wind models would behave if the dipole field were non-axisymmetric, in
particular, whether the outflow would originate from only those
regions of the inner disc edge where the stellar field was locally
strongest.

MUSICOS observations of SU Aur allowed this object to be monitored at
high spectral resolution continuously over 10 nights (Oliveira et al,
2000).  Profile variability of the principal Balmer lines and also Na
I and He I D3 (5876 \AA) was monitored. An approximate half
period shift was found between the variation of the red and blue wings of
H$\beta$, indicating outflow and infall are approximately in
antiphase. Smaller time lags were found between the He I line and the
redshifted component of H$\beta$, both supposedly formed close to the
stellar surface. The authors attribute this to an eclipse of a
twisted accretion stream. The twisting is then interpreted as being
caused by the differential rotation between the star and inner disc
creating toroidal field.

\begin{figure}
\psfig{{figure=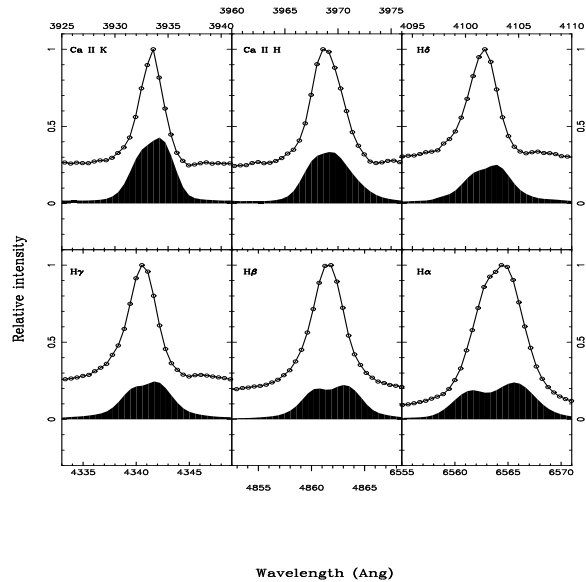,width=3.truein,height=3.truein}}
\caption{\label{variance} The standard deviation
across the line profiles (shaded)
compared to the average profiles (lines plus points). }
\end{figure}

Despite these tentative successes in understanding T~Tauri
environments in terms of an inclined dipole scenario, the possibility
remains that the spectral variations may be caused by non-rotational
effects. Variable accretion rates or magnetic reconnection in the
magnetosphere would be candidates to produce such variations. The
buildup of toroidal field by shear between disc and star would need to
be dissipated if the field geometry is to be maintained, which leads
us to expect some reconnection, with associated plasma heating and
hence variability. Meanwhile, a persistant problem in the channeled
accretion scenario is the loading of material onto the field lines
before it can fall onto the star. Simulations by Miller \& Stone
(1997) illustrate this for an aligned dipole geometry. The field lines
in the disc plane are forced to bow inwards by the pressure of
accretion through the disc, creating a local potential minimum from
which disc material cannot escape. Accretion through the disc proceeds
to build up material in this potential well, until the accretion
pressure overwhelms the field and the material accretes at the stellar
equator. This process could perhaps be circumvented in a non-aligned
dipole geometry, but the possibility of unsteady accretion, through
this or another mechanism, remains.  Examples of models based on
variable accretion include a ``beat frequency'' model proposed by
Smith et al (1995), which is analogous to that envisaged for low mass
X-ray binaries, to produce quasi-periodically varying accretion from
the inner disc. Ultchin et al (1997) suggested a model in which blobs
of disc material are dislodged from the disc by stellar magnetic field
and orbit the star, impacting either on the star or disc and giving
rise to variations.

\begin{figure*}
\vbox{
\hbox{
\psfig{{figure=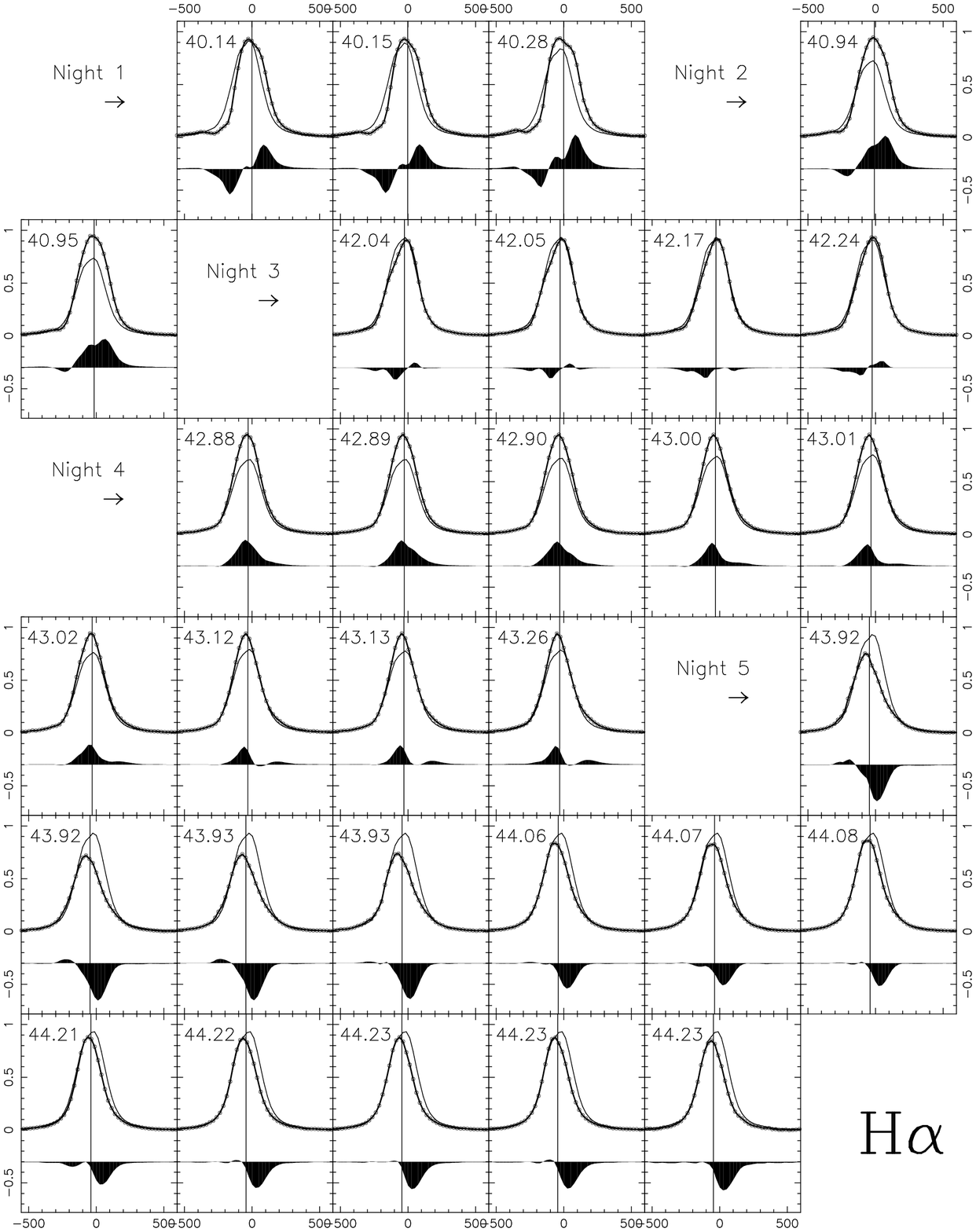,width=2.8truein,height=2.8truein}}
\hspace{0.2truein}
\psfig{{figure=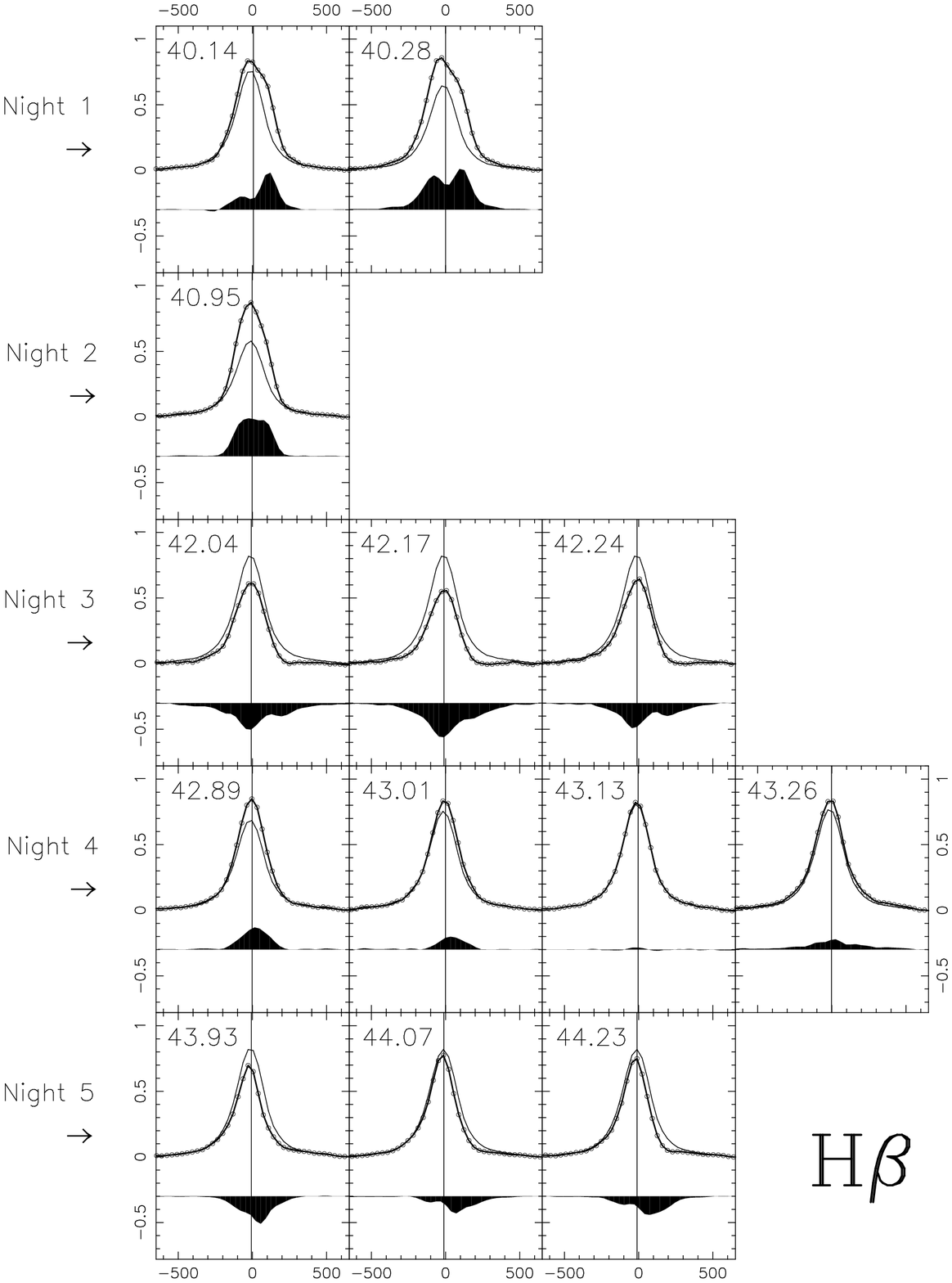,width=2.8truein,height=2.8truein}}
}
\vspace{0.2truein}
\hbox{
\psfig{{figure=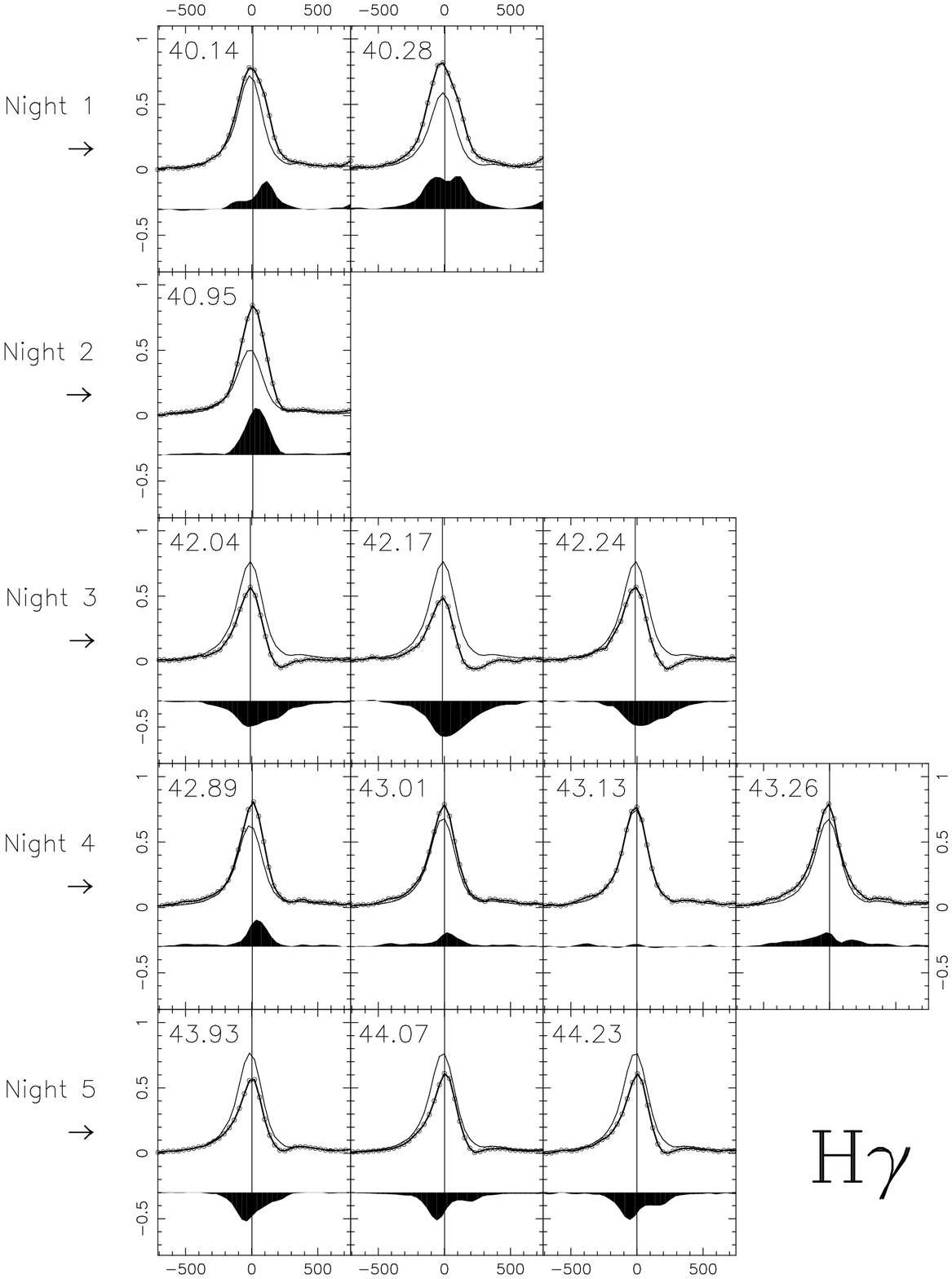,width=2.8truein,height=2.8truein}}
\hspace{0.2truein}
\psfig{{figure=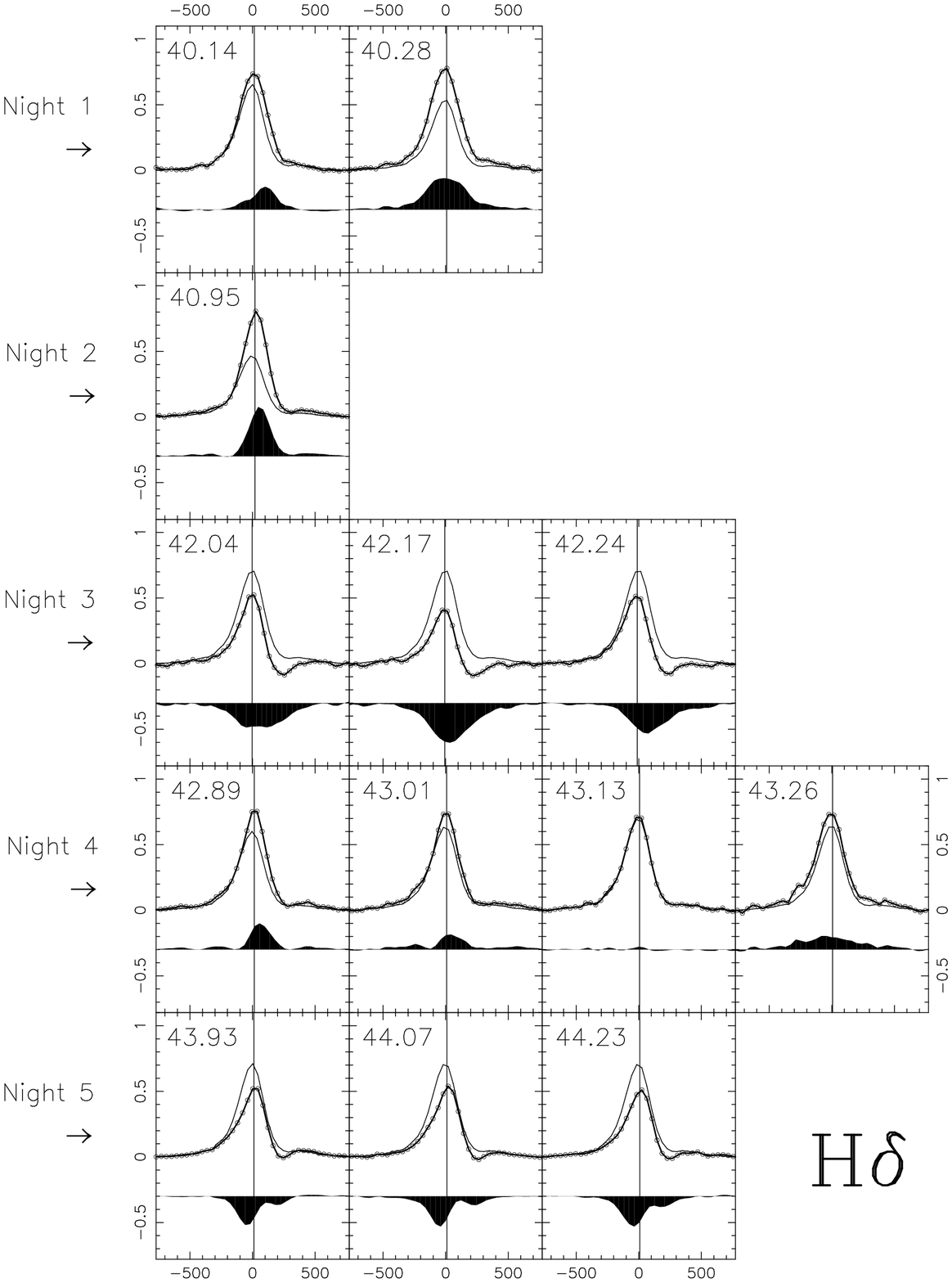,width=2.8truein,height=2.8truein}}
}
\vspace{0.2truein}
\hbox{
\psfig{{figure=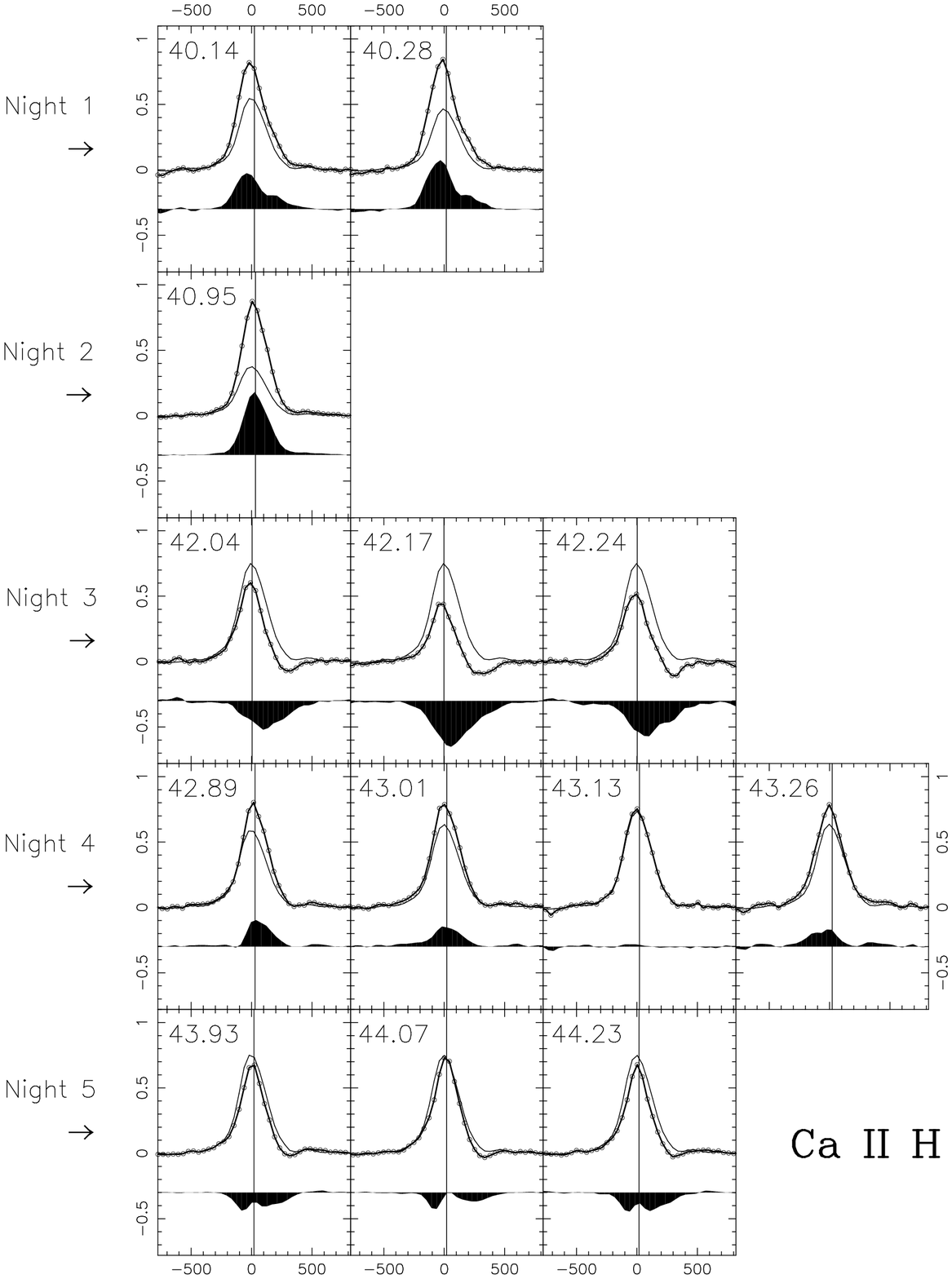,width=2.8truein,height=2.8truein}}
\hspace{0.2truein}
\psfig{{figure=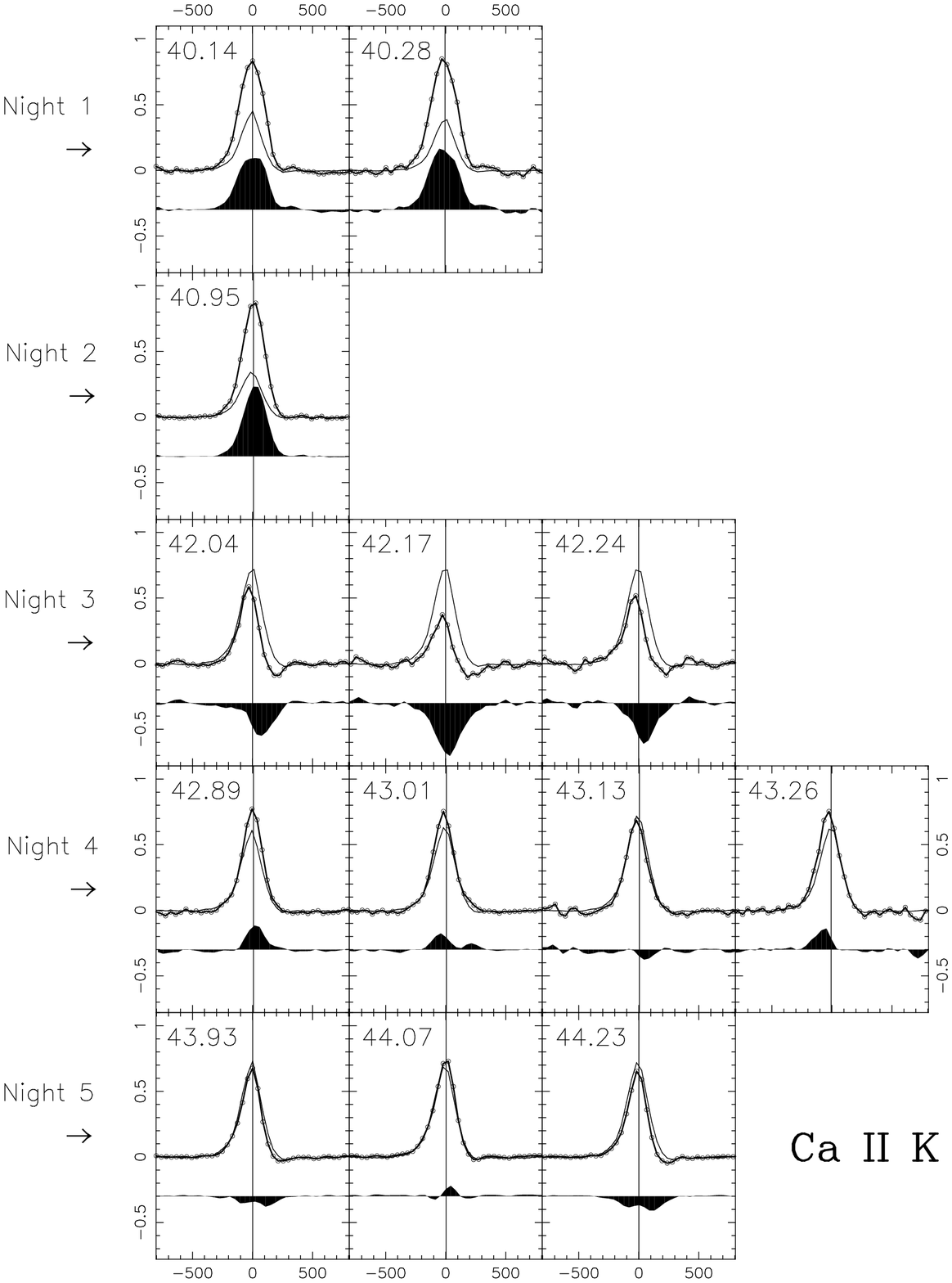,width=2.8truein,height=2.8truein}}
}
}
\caption{\label{spect} The profiles of the strongest lines over the
course of the observing run.  The continuum-normalised
spectrum is shown as a heavy solid line with points, the average
profile is shown as a light solid line, and the difference between
these (the residual emission) as a shaded area. The centroid of the
line is also shown in each panel (vertical line). The spectra have
been scaled vertically so that the maximum point of either the average
or programme spectrum is set at one and the continuum level is at
zero. The residuals have been displaced downwards by 0.3 for clarity.
For the lines in the blue spectrum, the spectra from each night are
shown on a separate line. For H$\alpha$, to save space, the various
nights data are shown consecutively and the separations between the
successive nights of the observing run are indicated. The observation date
(actually JD-2450900.0) is noted in each panel. }
\end{figure*}

The situation in the centre of T Tauri systems is therefore far from
clear. The cartoon picture of the star having an ordered dipole field,
which then channels smooth accretion flows onto the magnetic poles, is
almost certainly a crude simplification, although it seems that
elements of it are useful. The full picture is too complex for
detailed modelling to be a viable strategy. We must therefore
continue to glean clues to the morphology of the magnetosphere through
further observations.

In the current study, we present spectroscopic observations of the
CTTS VZ Cha over five consecutive nights, corresponding to two and a
half times the rotation period. On some nights, we obtained several
spectra and so we are able to comment on variations on timescales of
hours.  We observed extreme profile variability of several different
types, including redshifted absorption and emission, blue shifted
absorption and emission, and equivalent width variability. We describe this 
variability and attempt to decompose the profiles into distinct components.
We then examine the behaviour of the line components in 
the context of variable accretion and rotational modulation 
pictures.

\section{VZ Cha}

VZ Cha is a Classical T Tauri star with a K6 type photosphere.  It is
located in the Chamaeleon low mass star forming region at a distance
of approximately 160pc. The apparent V magnitude is 12.75 and highly
variable (Herbig \& Bell, 1988). The star is a strong H$\alpha$
emitter. The continuum veiling has not been previously measured, but
the lack of any absorption lines in our blue spectra suggests the star
is heavily veiled. Intermediate-resolution spectroscopy by Appenzeller
et al (1983) revealed redshifted absorption at H$\beta$.  Higher
resolution spectroscopy then established the presence of various
interesting spectroscopic phenomena, such as He II emission
(indicative of hot plasma), blue shifted emission at a velocity of
189 kms$^{-1}$ at H$\alpha$ and redshifted absorption in the Balmer
lines at velocities of around 240 kms$^{-1}$ (Krautter et al, 1990).
Drissen et al (1989) found linear polarization, indicating the
presence of a circumstellar disc. A photometric period of 2.56 days
was found by Batalha et al (1998), which is at odds with previously
reported photometric periods of around 7 days (Kappelman and Mauder
1981, Mauder and Sosna 1975) although it should be noted that these
last two studies were not based on sufficient data to derive
statistically significant periods, but rather found suggestive
evidence of periods in lightcurves when examined by eye.  A $K$-band
speckle survey for binarity by Ghez et al (1996) did not detect any
close companion, 
the limiting flux ratio 
being approximately 1 for a sky separation down to 0.1 arcsec (16AU at 
160pc).

\begin{figure}
\vbox{
\psfig{{figure=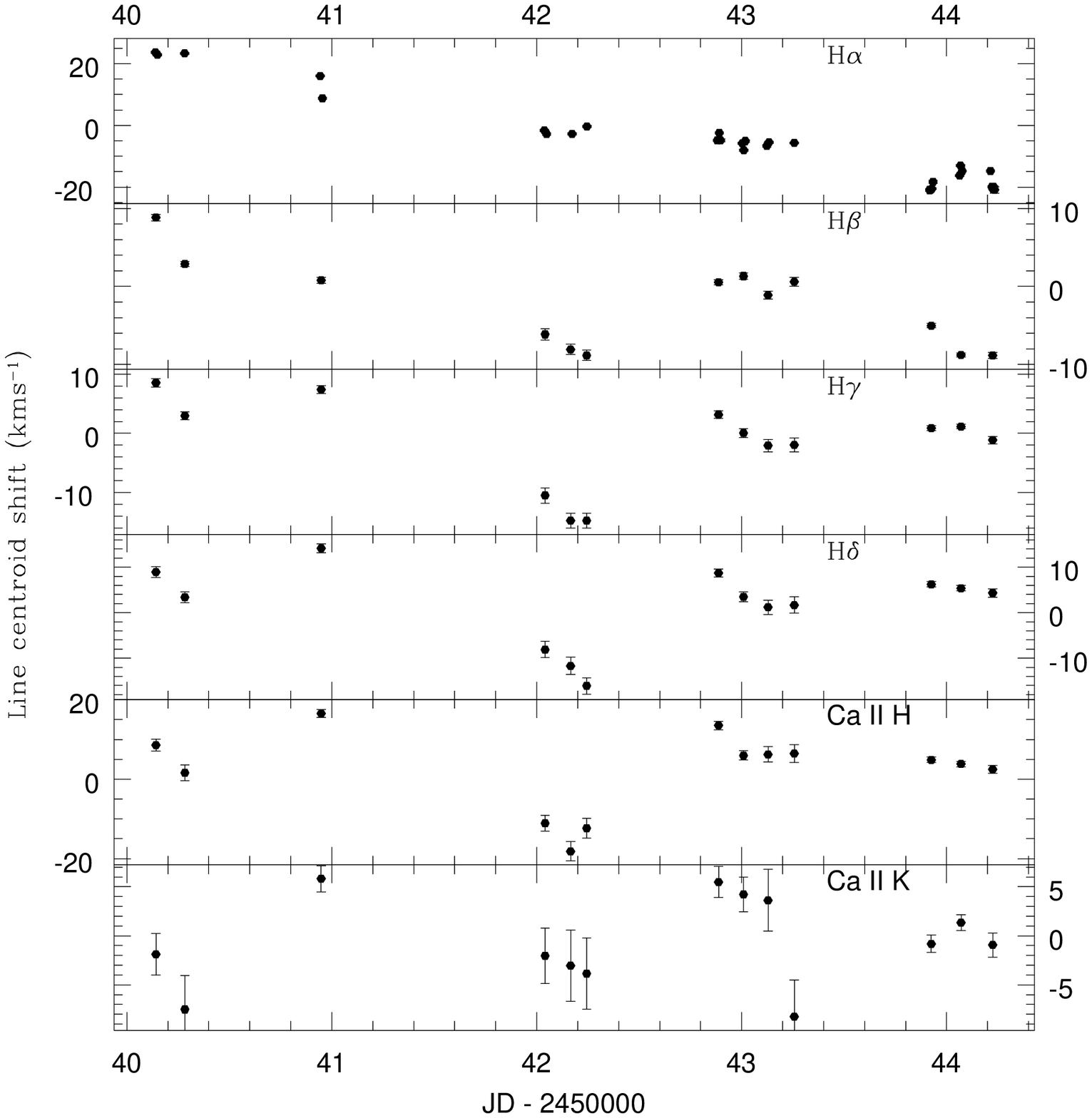,width=3.truein,height=3.truein}}
\psfig{{figure=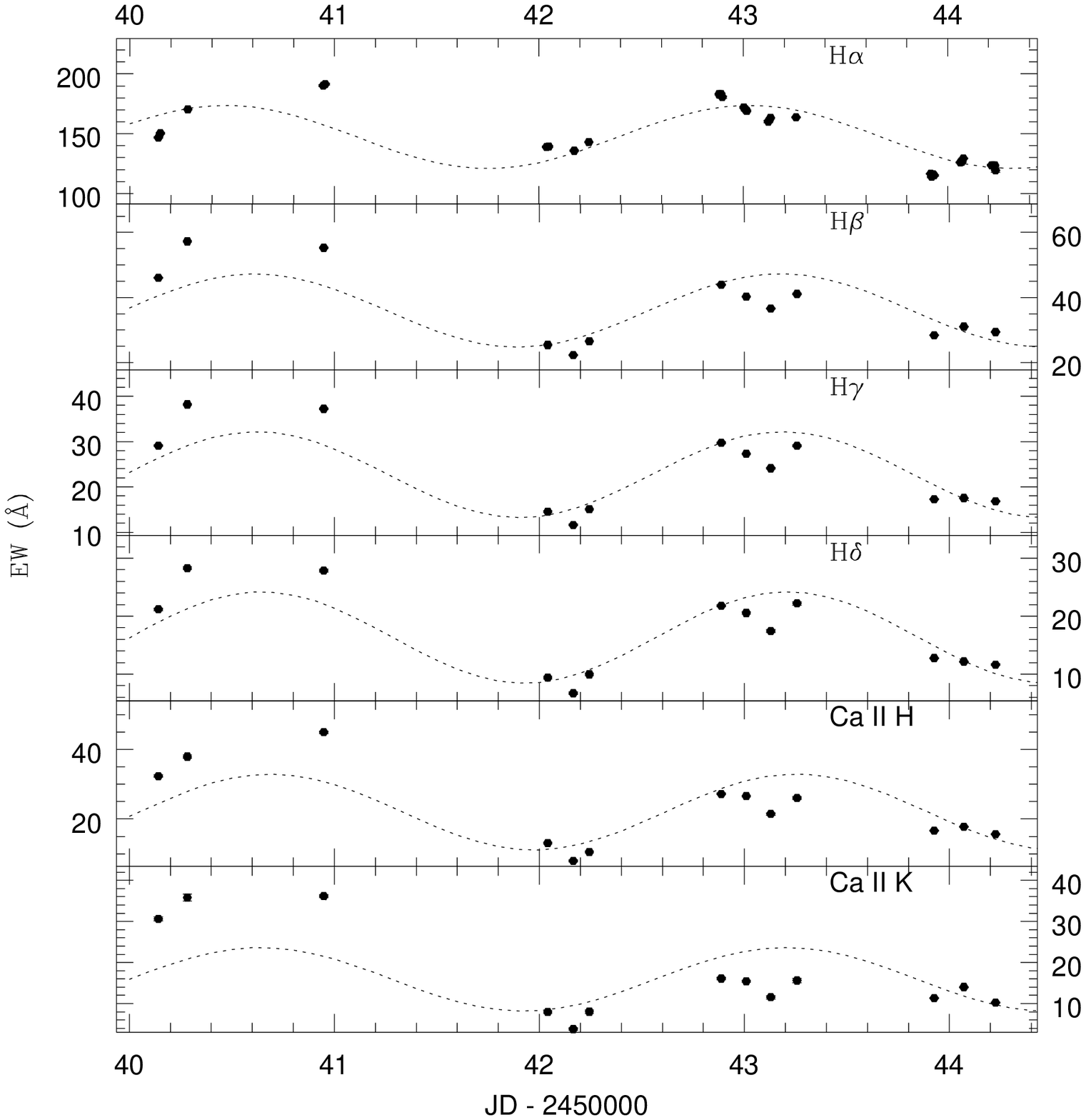,width=3.truein,height=3.truein}}
\psfig{{figure=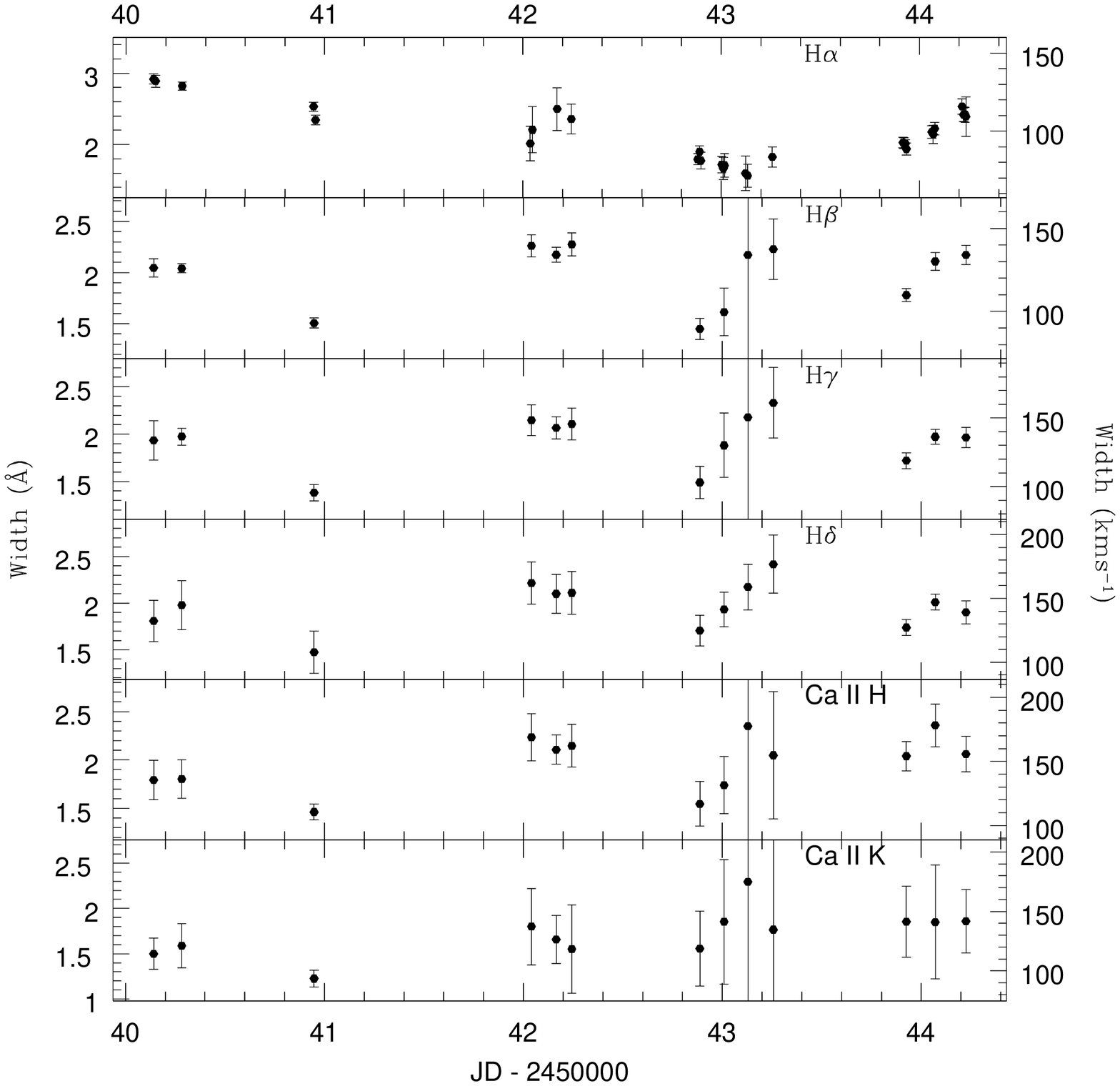,width=3.truein,height=3.truein}}
}
\caption{\label{centplot} The centroid shift of the spectral line (top), 
the equivalent widths (middle) and the residual widths (bottom) plotted against
time. Error bars in each case are calculated from the flux errors
determined in the data reduction. Positive centroid offsets indicate
the line centroid was redshifted. Positive equivalent widths denote
emission lines. A sine curve has been overplotted in the middle
panel.}
\end{figure}

\begin{figure}
\vbox{
\psfig{{figure=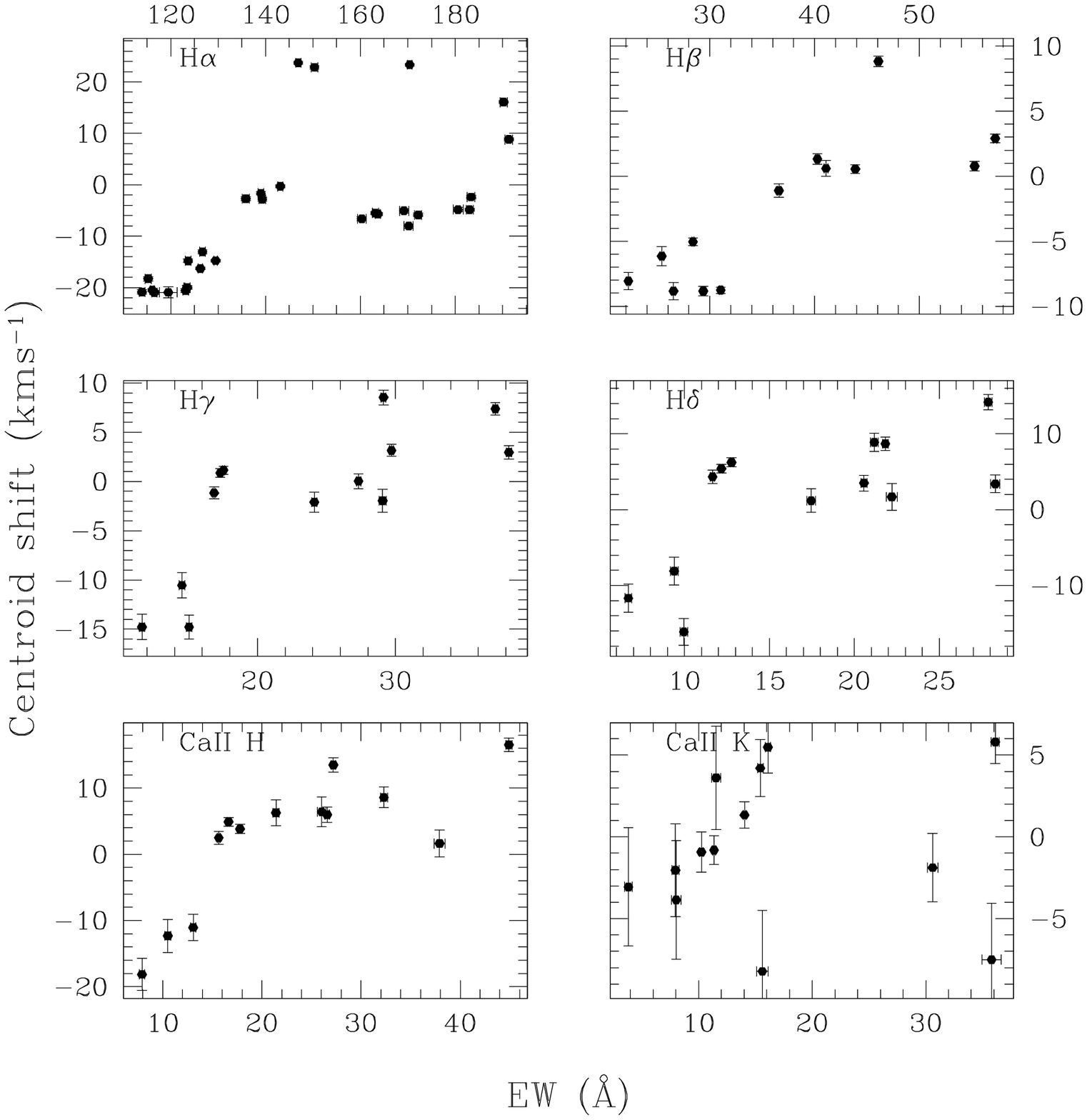,width=3.truein,height=3.truein}}
\psfig{{figure=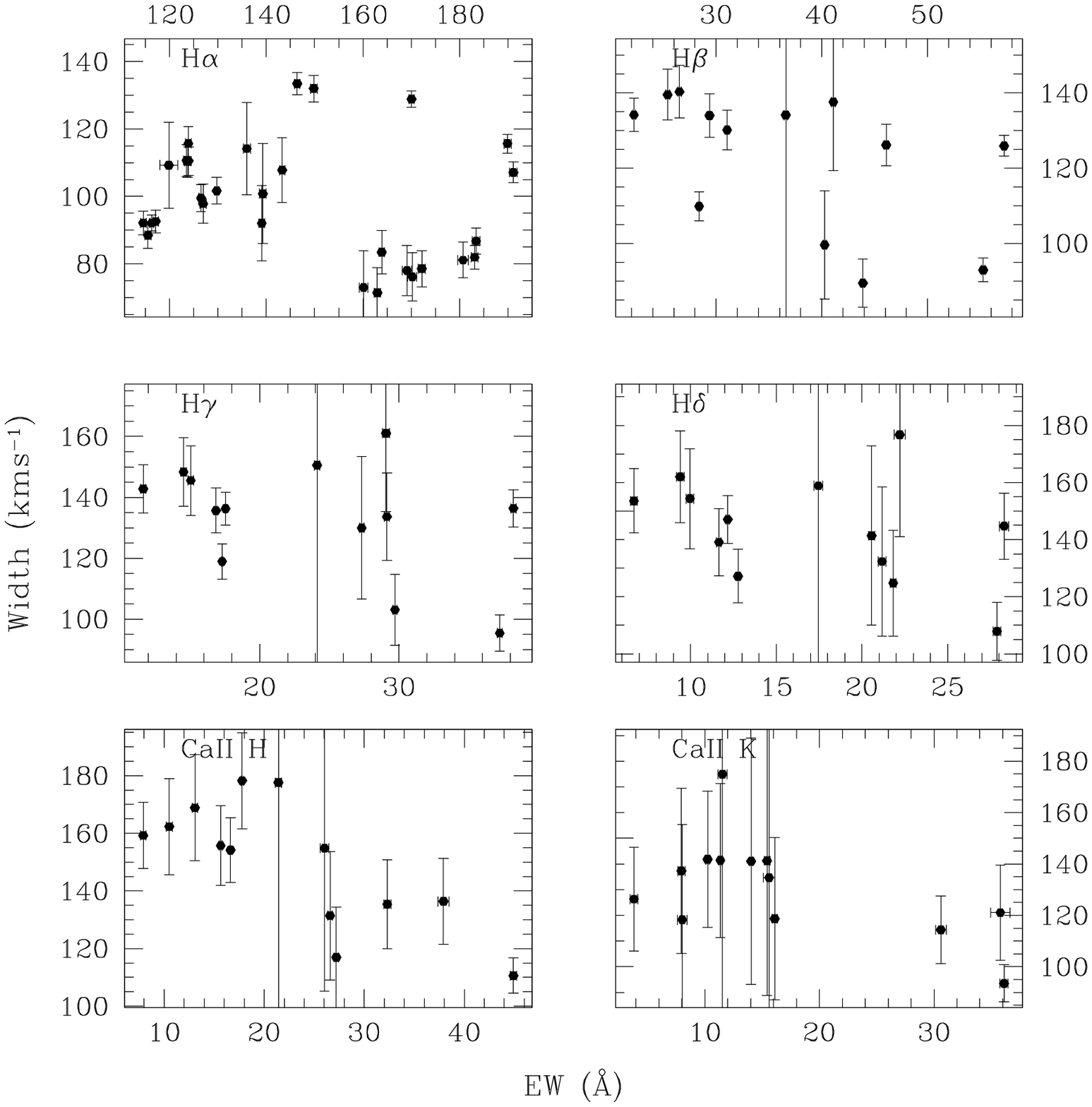,width=3.truein,height=3.truein}}
\psfig{{figure=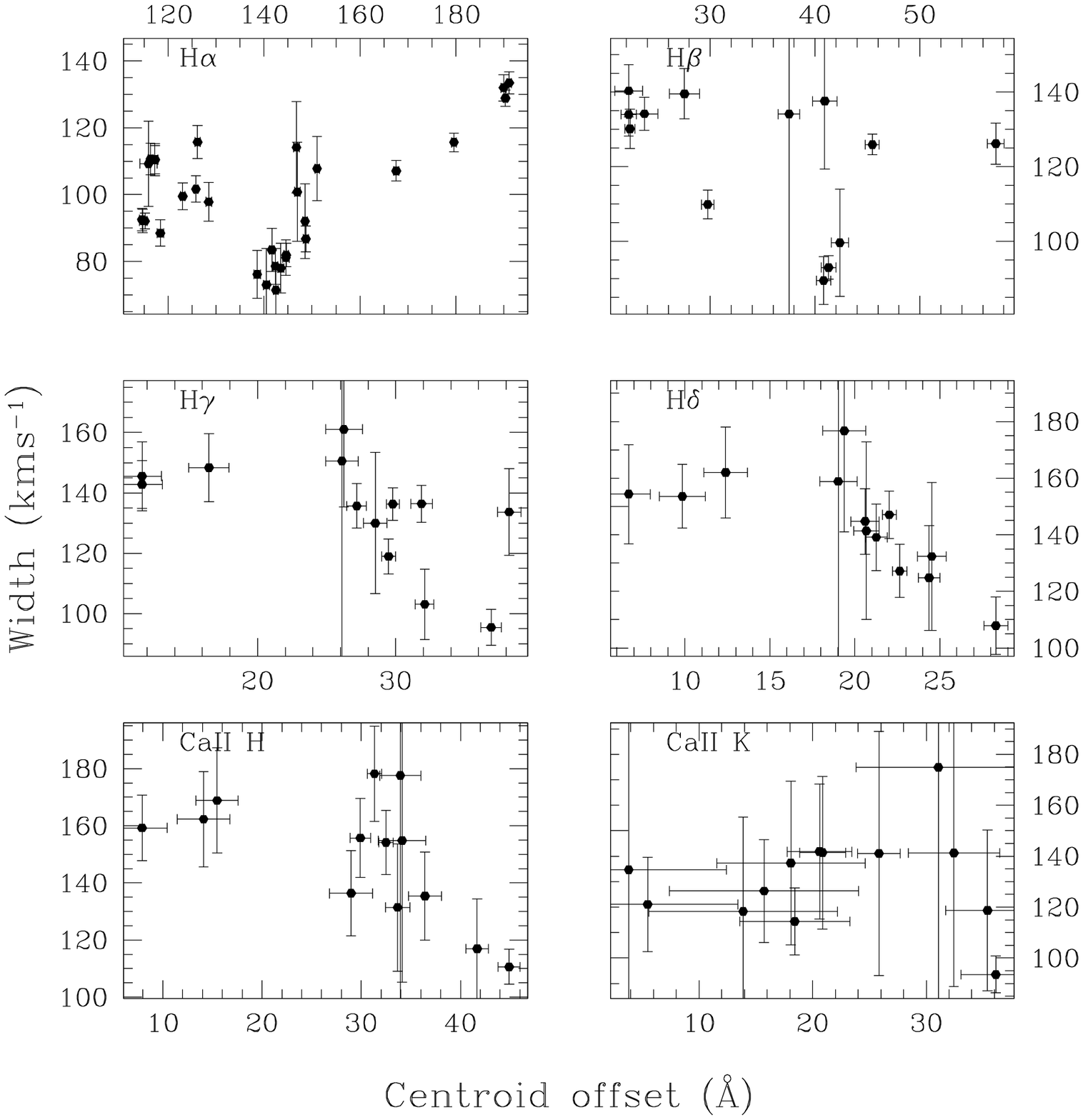,width=3.truein,height=3.truein}}
}
\caption{\label{centvew} Line centroid shift versus equivalent width
(top panel), residual width versus equivalent width (middle), and residual 
width versus centroid offset (bottom panel). }
\end{figure}

\section{The observations}

The observations were carried out between 6/7 and 10/11 May 1998 with
the 2.3m ANU telescope, at the Siding Spring observatory in New South
Wales, Australia.  The instrument used was the dual beam
spectrograph. Gratings of 1200 lines/mm were employed in each arm. This
allowed us to observe (at first order) the spectral region from
approximately 3900\AA\ to 4880\AA\ in the blue, and from approximately
5680\AA\ to 6640\AA\ in the red arm.  Because our sensitivity was
greater in the red, we were able to obtain several exposures of the
red spectrum for every blue exposure.  The data were reduced using
standard IRAF\footnote{IRAF is distributed by the National Optical
Astronomy Observatories, which are operated by AURA under co-operative
agreement with the National Science Foundation.} routines. The
dispersion was approximately 0.5\AA\ per pixel in each arm. The FWHM
of a typical unblended arc line was found to be approximately 2.0\AA.
The wavelength scale was calibrated by reference to arc spectra
observed during the run. The RMS of the fits to the arc-lines was
found to range between 0.08 and 0.15\AA.

No photometric standards were observed, since the observations were
not carried out in photometric conditions. The continuum was therefore
removed by fitting with a Chebyshev polynomial and dividing it out.

\section{Analysis}

Various types of profile variability were seen, including redshifted
and blueshifted emission and redshifted absorption. This variability
occured predominantly on night-to-night timescales. The clearest
variation was the appearance of Inverse P Cygni (IPC) type profiles on
nights three and five.  The pixel by pixel standard deviation across
the profiles of the six strongest lines was computed and is shown in
Figure~\ref{variance}. For comparison an average profile constructed
by taking the mean of all the available spectra, excluding the minimum
and maximum at each pixel to prevent unwanted effects due to outliers,
is also shown.  For the Balmer lines at least, the variability seems
to be divided into two main components, one in the blue wing and one
in the red wing. We estimate the red component to lie at a velocity
relative to the line centre (measured from the average profile) of
approximately 70kms$^{-1}$, a value which remains more or less
constant from H$\alpha$ through to H$\delta$.  The varying component
in the blue wing steadily increases in velocity shift from around
50kms$^{-1}$ in the case of H$\delta$ up to 130kms$^{-1}$ for
H$\alpha$.  The shape of the average H$\alpha$ profile appears to have
a blue hump, indicating a wind component which is present to some
extent most of the time.  To further quantify this variability, we
subtracted the average spectrum from each of the programme spectra to
obtain a measure of the variable residual emission at different times.

\subsection{Description of line profile variations} 

The spectra, average profiles, and residuals are plotted in
Figure~\ref{spect}. Variations on both sides of the profiles are
seen. Below, we describe the
features seen in the residual profiles. We note first that some care
must be taken when describing lack of emission in a profile relative
to the average profile. Where the actual profile dips below the
continuum level (which in Figure~\ref{spect} is at the zero flux
level), we can be sure that absorption is occuring. However, where
there is a lack of flux relative to the continuum, but the spectrum
does not dip below the continuum, we may be seeing absorption or
merely a comparative lack of emission. Hereafter we will refer to the
former case as {\em absorption} and to the latter as a {\em
deficit} of emission, in contrast to {\em excess} emission.

Classical T Tauri spectra typically contain a strong excess continuum
which veils the stellar spectrum. This veiling continuum is variable
and this will cause variations in the equivalent widths of the
emission lines. Attempts were made to measure the veiling variations
amongst our red spectra by measuring the variation of absorption lines
(the blue spectra contained no significant absorption lines at
all). These lines would be assumed to be purely photospheric. The most
prominent absorption features available were the Na D lines at about
5890\AA,. These revealed variations which were apparently slightly
anticorrelated with the equivalent width variations of the Balmer
lines. This would be expected from variable veiling, but also from
filling-in of the absorption lines by extra line flux. Since the Na D
lines have been seen to show some non-photospheric emission flux in
other CTTS (Edwards et al 1994), we regard our veiling estimates as
unreliable and have not included them. Variations in the veiling will
of course produce variations in the emission line strength across the
whole profile.

The higher Balmer lines (from H$\beta$ onwards) and Ca II H and K show
clear evidence of redshifted {\em and} blueshifted excess emission
early in the run (nights one and two).  H$\alpha$ by contrast shows
excess red emission and a deficit on the blue side of the line.  A
central emission component appears and grows in the H$\alpha$ profile
over nights one and two, whilst the blue deficit diminishes. By night
three, the excess in the higher Balmer and Ca II lines has given way
to a general emission deficit across the profile, with the main part
in the red wing. H$\alpha$ meanwhile shows small deviations from the
average profile, with the red side slightly in excess and the blue
side in deficit. The H$\alpha$ profiles on night three appear by
eye to be skewed towards the red. On night three, 
at H$\gamma$, H$\delta$
and the Ca II lines, the redshifted features dip below
the continuum level and can therefore be classified as true absorption
features. It
should be noted here that Ca II H at 3969.6\AA\ will be blended with
H$\epsilon$ at 3971.2\AA, so that behaviour in the red wing of Ca II H
which differs from that observed at Ca II K is possibly due to
variations in H$\epsilon$.  The profiles of these lines return to an
emission excess by night four, although there is less excess emission
than formerly. There is some evidence in the higher Balmer lines that
the narrow central excess at the beginning of the night is replaced by
a very broad excess component at JD=43.26.  On night four H$\alpha$ 
also shows excess emission, preferentially on the blue side with a
long tail into the red, and with a steady decrease in the emission
near the profile centre.  By the fifth night, deficits are again seen
for all lines (with the possible exception of Ca II K). This time the
main component is located in the blue wing for H$\gamma$, H$\delta$
and Ca II H. At H$\beta$ the deficit is more pronounced on the red
side of the line and at H$\alpha$ the deficit is entirely on the red
side.

\subsection{Time series behaviour}

We computed the centroids of the lines over a wavelength range centred
on the centroid of the average profile. The line centroid is 
\begin{equation}
\lambda_c=\frac{\Sigma |f_p| \lambda_p }{\Sigma |f_p|},
\label{centroid}
\end{equation}
where $f_p$ are the fluxes and $\lambda_p$ the wavelengths at each
pixel.  A major consideration in this calculation is the range over
which the sum is taken. We chose a range of 12 pixels which
ensures the entire profile is covered.  The centroid of the average
profile was chosen as the centre of this range as it is the most
stable estimate of the true line position. Since the pixels
of the average profile are interpolated to the same wavelength values
as the spectrum before subtraction, the centroid of the average
profile is not exactly the same for each spectrum. The variations from
spectrum to spectrum due to interpolation are typically of the order
of thousandths of an \AA ngstrom, and so not significant compared to
the centroid variations caused by spectral variability. 

An additional problem arises because the value of the average profile
centroid will typically lie between pixels in the spectra. To counter
this, we introduced a weighting factor to the first and last terms in
each sum in Equation~\ref{centroid}. These fluxes were weighted 
in inverse proportion to the proximity of the average spectrum centroid.
For example, if the interpixel spacing
is $\delta \lambda$, and the average spectrum centroid lay
0.1$\delta \lambda$ from pixel $n$ and 0.9$\delta \lambda$ from pixel
$n+1$, we would assign a weight of 0.9 to the {\em first} pixel in each sum
in Equation~\ref{centroid} and a weight of 0.1 to the {\em last}. 

The line centroid positions are marked with vertical lines in
Figure~\ref{spect}. We also computed the equivalent widths of the
lines.  Both the centroids and equivalent widths are shown in time
series form in Figure~\ref{centplot}.  Also shown in the bottom panel
is the root mean square of the dispersion of the residual about the
line centre. This is defined as
\begin{equation}
W=\sqrt{\frac{\Sigma |f_p| (\lambda_p - \lambda_c )^2}{\Sigma |f_p|}},
\end{equation}
where $\lambda_p$ is the wavelength and $f_p$ the residual value at
each pixel and $\lambda_c$ is the line centroid. We will refer to this
quantity as the 'width' of the residual. It is insensitive to
differences between absorption and emission, and also relatively
insensitive to changes in the veiling, and is intended to serve as a
measure of the velocity dispersion of material causing changes in the
line profile. The ordinate in the lower panel of Figure~\ref{centplot}
has been labelled in wavelength units (on the left side) and in
equivalent velocity (on the right side).

From Figure~\ref{centplot} we can make a number of observations.  The
centroid of H$\alpha$ moves from red to blue in a gradual trend over
the entire observing run with no discernable shorter timescale
variations. This reflects the development of deep redshifted
absorption over the course of the observations (cf
Figure~\ref{spect}).  The other lines show more complicated
behaviour. The possibility of periodic effects is discussed in a
separate subsection (\ref{periods}). There is a tendency to exhibit a
blueshifted centroid on night three, indicating {\em redshifted
absorption} which is also clear in Figure~\ref{spect}. There is also a
tendency to show redshifted centroids on night one, indicative of {\em
excess redshifted emission}, and also on night five, where they seem
to indicate instead a lack of blue flux. 

The lower panel of Figure~\ref{centplot} shows that the residual
widths are generally highest on night three, when the redshifted
absorption was strongest for most lines.  The exceptions are
H$\alpha$ and Ca II K.  The residuals show the least dispersion about the
line centre on night two for all the lines except H$\alpha$.  The
widening of the residual profiles for most lines on night four is
visible in Figure~\ref{spect} and is clear for most of the lines in 
Figure~\ref{centplot}.  

We have plotted the variables against one another in
Figure~\ref{centvew} and searched for correlations between them.
Correlation coefficients and resulting significances were computed for
each pair of quantities.  The results appear in
Table~\ref{corrtable}. It is clear that strong correlations exist
between equivalent width and centroid shift for most lines. This
indicates that the equivalent width variability is primarily caused by
variation in the red wing of the line, since a deficit in the red will
push the centroid to the blue side (a negative centroid shift in our
scheme), whilst red emission will drag the centroid to the red (a
positive shift). This is consistent with the standard deviations shown
in Figure~\ref{variance}. The exception is Ca II K, for which the
correlation is not significant.

The correlations between equivalent width and residual width and
between line centroid and residual width are generally less
significant. In the case of equivalent width versus residual width, a
positive or negative correlation would indicate that the varying
components (whether emission or absorption) occur in the line
wing. Variation in both wings simultaneously would cause the largest
increase in residual width. Variations in either wing, either with
flux excess or flux deficit relative to the average spectrum, will cause the
residual width to increase. In the equivalent width
versus residual width plots, H$\beta$ and Ca II H show slightly
significant anticorrelations.  The fact that these correlations are
negative may indicate that the dominant transient features are
absorption features.

Correlations between the residual width and line centroid shift
probably indicate activity in the wings of the lines. Significant
anticorrelations (by the non-parametric test) are seen for H$\gamma$
and H$\delta$.  This may indicate red flux deficits causing the line
centroid to shift to the blue (negative) whilst increasing the
residual width. These correlations are not immediately evident to the
eye, however, and the parametric correlation coefficients are modest,
perhaps indicating that the relationship is complicated by the appearance of
red excess on night one or blue deficit features on night five.

To further characterise the time series behaviour of the centroids,
the lines were split in two, taking the middle point to be the peak of
the average spectrum at each line. The total residual on either side
of the line centre was then determined. This tactic was intended
primarily as a search for an anticorrelation between red shifted
absorption and blue shifted emission, as reported by previous authors
(Johns \& Basri, 1995, Oliveira et al 2000).
\begin{figure}
\psfig{{figure=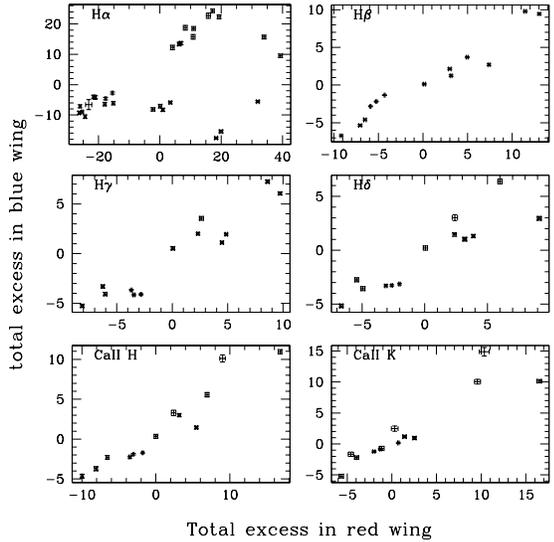,width=3.truein,height=3.truein}}
\caption{\label{redblue} The total emission
in the red wing versus the total emission in the blue wing of 
each line. The lines are split into two halves with the centre at the
position of the peak of the average profile after rebinning onto the
wavelength scale of the spectrum.  }
\end{figure}
Figure~\ref{redblue} shows correlation plots of the total red residual
versus the total blue residual. The red and blue sides of most of the
lines vary together - as more red flux is present, so more blue flux
is also observed. This situation would be expected if the line
variations are driven mostly by variations in the veiling continuum.
Only at H$\alpha$ is this correlation not always seen. The correlation
coefficients in Table~\ref{corrtable} show that the correlations of
red and blue residuals are significant at levels of 99.9\% for all the
lines except H$\alpha$.

\begin{table*}
\begin{center}
\begin{tabular}{|c|ccc|ccc|ccc|ccc|} \hline 
            &   \multicolumn{12}{|c|}{Quantities correlated} \\ \hline
            &   \multicolumn{3}{|c|}{EW vs. line centroid} & \multicolumn{3}{|c|}{EW vs. line width} &\multicolumn{3}{|c|}{Centroid vs. width} & \multicolumn{3}{|c|}{Red vs. blue residuals} \\
            &    \multicolumn{3}{|c|}{Figure~\ref{centvew} top}&  \multicolumn{3}{|c|}{Figure~\ref{centvew} middle}&  \multicolumn{3}{|c|}{Figure~\ref{centvew} bottom}& \multicolumn{3}{|c|}{Figure~\ref{redblue}} \\ \hline
Line          & r$_{sp}$ & P      & r$_c$     & r$_{sp}$ &    P  &  r$_c$   & r$_{sp}$ &    P   &  r$_c$  & r$_{sp}$ &    P&  r$_c$   \\ \hline
H$\alpha$     & 0.75   &$>$99.9\% &   0.65    & -0.25 &   80\%   & -0.24    &  0.20 &  50\%     &  0.42   & 0.27   & 80\%  &  0.31  \\  
H$\beta$      & 0.81   &$>$99.9\% &   0.81    & -0.63 &   98\%   & -0.51    & -0.53 &  90\%     & -0.40   & 0.99   & $>$99.9\% & 0.97 \\
H$\gamma$     & 0.81   &$>$99.9\% &   0.75    & -0.41 &   80\%   & -0.41    & -0.73 &  99.5\%   & -0.56   & 0.87   & $>$99.9\% & 0.92 \\    
H$\delta$     & 0.56   & 95\%     &   0.69    & -0.38 &   80\%   & -0.37    & -0.85 &$>$99.9\%  & -0.66   & 0.87   & $>$99.9\% & 0.85    \\  
Ca II H       & 0.77   & 99.8\%   &   0.77    & -0.72 &   99\%   & -0.75    & -0.62 &  95\%     & -0.58   & 0.96   & $>$99.9\% & 0.92   \\      
Ca II K       & 0.29   & 50\%     &   0.04    & -0.46 &   80\%   & -0.57    &  0.10 &  20\%     &  0.04   & 0.93   & $>$99.9\% & 0.86   \\ \hline
\end{tabular}
\caption{\label{corrtable} Correlations between various
quantities. The standard parametric correlation coefficient (r$_c$),
Spearman rank order correlation coefficient (r$_{sp}$), and the
significance of the correlation (P) are shown for each line. The
significance level is found by a two-tailed Student's $t$ test based on
the Spearman correlation coefficient. No attempt was made to test the
significance of the parametric correlation.}
\end{center}
\end{table*}

\subsection{Periodic effects}
\label{periods}

Our time series' are too short and poorly sampled to make a search for
periodicities in various features, but we can comment on the
consistency of the data with various supposed rotational effects.  As
already noted, the photometric period of VZ Cha has been measured to
be 2.56 days, and the implication is that this represents the stellar
rotation period.  This photometric variation could have been caused by
the presence of dark photospheric spots, or by bright regions such as
the accretion shocks at the magnetic poles. In the latter case we
would expect to see the emission lines, which supposedly form in the
accretion stream, modulated with the same period.  If the photometric
variation is caused by dark spots, then the line emission could be
modulated with the same period or with a higher harmonic of it if
there are multiple streams. In principle it is also possible that
there are more spots than streams, so that the 2.56 day photometric
period is a high harmonic of the true period, and the line period
would then be longer. As well as modulation of the line strengths, we
might expect to see modulation of the components at various
characteristic velocities as the non-axisymmetric velocity field of
the accreting material rotates. This could lead to periodic behaviour
in the line centroids. It should be noted that any rotational
modulation would not necessarily be sinusoidal in shape.

To illustrate the discussion of periodicity for the equivalent widths,
we show overplotted sine curves in the right-hand panel of
Figure~\ref{centplot}.  These curves are constrained to have a 2.56
day period, as found by Batalha et al. The amplitude, phase and
y-offset were left as free parameters. The reduced $\chi^{2}$ of these
fits is very large in all cases, which would be expected from inspection of
Figure~\ref{centplot}. It can be seen that the underlying
night-to-night modulation of the equivalent widths roughly matches the
expectation from the sine curve. We recomputed $\chi^{2}$ for the
night-to-night variations, using nightly means of the equivalent width
and Julian date and determining standard deviations of each nightly
distribution separately. These are then interpreted as characterising
the supposed short-timescale variations superimposed on the
night-to-night variations.  Night two was excluded for all the lines
except H$\alpha$, as there was only one point available and so it was
not possible to estimate the short timescale variation.

With this procedure, the best fits are found to be H$\delta$ and
H$\gamma$ with reduced $\chi^{2}$ of 3.4 and 4.1 respectively.
H$\beta$ has a reduced $\chi^{2}$ of 7, and Ca II H and K have
$\chi^{2}=14$ and $30$. H$\alpha$ still has a large $\chi^{2}$ of
several hundred indicating a poor fit, but for H$\alpha$, the night
two data is retained. Including the point from night two for the other
lines, and taking the measurement uncertainty in this as $\sigma$, the
reduced $\chi^{2}$ increases for all the lines, the best still being
H$\delta$ with a reduced $\chi^{2}$ of 8.7. Finally we refitted
H$\alpha$, using the night by night data and $\sigma$'s, and obtained a
fit with reduced $\chi^{2}$ of a little over 17. This represents the
best possible sine curve fit to the H$\alpha$ equivalent width time
series.

Similar fits were made to the centroid time series'.  These fits were
in general less successful at reproducing the night to night behaviour
than the fits to the equivalent width time series'. It can be seen
from Figure~\ref{centplot} that only H$\beta$ and to a lesser extent
Ca II K display the night to night variations which might be
consistent with a period of around 2.5 days. Finally we note that the
behaviour of the H$\alpha$ centroid is entirely inconsistent with a
periodic effect with a period shorter than the data set. If the
H$\alpha$ centroid behaviour is periodic, it must have a period of at
least 10 days.

\section{Discussion}

We concluded in Section~\ref{periods} that whilst the night-to-night
equivalent width variations of all the lines except H$\alpha$ might in
part be caused by a periodic modulation with a 2.56 day period, the
detailed behaviour of the equivalent width variations is inconsistent
with a sine wave variation. Furthermore, the amplitude of the
variation has changed between the two maxima in our dataset. This
indicates that there must be some other variation of the accretion
system superimposed on any rotational effect.

Could we reject the idea of rotating streams altogether and rely only
on variable accretion to explain the variations observed?  It is of
course possible to construct models based on varying accretion alone
which could explain the variations, but such explanations are in our
view unlikely. Many features in the profiles are seen to persist for
timescales of at least one night (for example redshifted absorption on
night three) or even two nights (for example the redshifted excess
emission on nights one and two). Since the infall time from the inner
disc edge is likely to be of order twelve hours (e.g. Paatz \&
Camenzind, 1991), we would expect spectral variations due to accretion
of distinct blobs to occur on timescales shorter than this. The
profile activity therefore suggests that accretion occurs in streams
which are stable over timescales longer than the actual infall time
and comparable to the rotation period.  It therefore seems likely that
the observed variations are indeed partly due to rotational
modulation.

Two episodes of absorption in the red wings of the lines are seen. The
first occurs during night three, predominantly affects H$\gamma$,
H$\delta$~and Ca II H and K, and causes genuine absorption
features which dip below the continuum. A second episode begins on night
four and affects H$\alpha$ strongly, persisting into night five when
the effect on H$\alpha$ diminishes but the feature is seen at
the higher Balmer lines as well.

The H$\alpha$ activity is in several cases inconsistent with the
activity of the other lines.  A similar situation was noted by
Johns-Krull \& Basri (1997) in the case of DF Tau. It seems H$\alpha$
must mostly originate in a more extended region than the other
lines. This is also suggested by the decoupled behaviour of the red
and blue wings of H$\alpha$ compared to the other lines, and is
expected due to the higher optical depth of H$\alpha$ compared to the
other Balmer lines. Nevertheless, H$\alpha$ does show a lack of flux
in the red wing on night five, when the other lines undergo a weaker
redshifted-absorption episode.  In this instance, the behaviour seems
to be more strongly coupled.

The night five observations are at a slightly earlier phase in the
rotation cycle than the night three observations. It is possible
therefore that we are seeing a twisted accretion stream, as suggested
for SU Aur by Oliveira et al (2000). The H$\alpha$, originating mostly
from the outer parts of the flow, is affected by the outer stream
passing through the line of sight. This would be the effect that we observe
on night five.  A fraction of a rotation later, the inner part of the
stream passes through the line of sight, and the optically thinner
higher Balmer lines show redshifted absorption features (as on night three).

Alternatively, it is possible that there are two accretion streams
causing the variations (the inclined dipole scenario). The 2.56 day
period could then be one half the true rotation period of the system,
and the events on night three and night five would represent the two
separate streams passing through the line of sight. A non-equatorial
observer's line of sight would intersect different parts of each
stream, and so the variation caused by each stream would show
different characterisitics. In particular, the inner part of one
stream would pass through the observer's line of sight, whilst only
the outer regions of the other may do so. This could lead to strong
obscuration of the higher Balmer lines by rapidly moving inner
material in the first stream, followed by effects seen predominantly
at H$\alpha$ which originates from the outer regions of the second
stream. It must be remembered that such a scenario requires that an
observer viewing from the northern hemisphere should have an
unobstructed view of at least part of the southern hemisphere
accretion stream, or vice versa.  We estimate that an observer viewing
from 11$^o$ above the disc plane, with the disc truncated at 5R$_*$
and fully transparent inside that orbit, should be able to view a
southern hemisphere accretion shock at an inclination of 10$^o$ or
more from the rotation axis of the star. Of course, such an observer
would view the inner stream region from almost side on, and so would
not see a large component of the infall velocity close to the stellar
surface.  Such an observer might see a large component of the infall
velocity from the central portion of a curved stream, and this
velocity could still be considerable (of order 100kms$^{-1}$ or more,
see again Paatz \& Camenzind, 1991). 

In either case, we would expect the highest velocities to be
associated with the inner part of the stream, and to be seen on night three.
This is borne out by the lower two panels of
Figure~\ref{centplot}, where the largest line centroid shifts are seen
on night three, and also the largest residual widths. The large
residual widths may indicate the phases where we see the largest
dispersion of velocities in the line of sight.  This is consistent
with the idea that we are looking down an accretion stream during the
redshifted absorption episode on night three. H$\alpha$ shows some
signs of behaving similarly to the other lines on night three, as the
residual widths increase slightly against the general downward trend
from night one to night four. Apparently, H$\alpha$ is affected only
slightly by an event which causes the high velocity dispersion
variations in the other lines. This may be the passage through the
line of sight of the inner part of an accretion stream, but with the
outer part of the stream not passing directly through the line of
sight. We favour this type of geometric explanation rather than an
explanation based purely on optical depth, since we see from night
five that H$\alpha$ can be strongly affected by high velocity receding
material.

\section{Summary}

Our data show that VZ Cha is an interesting object for future
study. The line profiles show definite redshifted absorption features
and other transient features which vary on timescales similar to the
rotation timescale. We found that the variation occurred in the red
wing more than in the blue wing of the line. The data we present here
is not sufficiently well sampled to rigorously examine the hypothesis
of periodicity for the line equivalent widths or profiles, but we can
assess the viability of a rotational picture for the activity of VZ
Cha. We find that the night-to-night variations in equivalent width
are to some extent consistent with a 2.56 day periodicity, but that
other variations, such as variable accretion or reconnection, are
necessary to account fully for the behaviour. It is unlikely that rotational
effects are entirely unimportant as the observed lifetimes of varying
profile components are longer than or comparable to the infall time
from the inner disc, and also comprise a significant fraction of the
rotation period. We found that the red and blue variations, probably
representing accretion and outflow respectively, were correlated for
all lines except at H$\alpha$, where there were episodes of
anticorrelated behaviour. This suggests that H$\alpha$ originates in a
more extended region than the higher Balmer lines and samples parts of
the accretion and outflow streams which are affected differently by
the system's rotation.

\begin{acknowledgements}

We thank the staff at SSO and ANU for providing the facilities on
which this paper is based. We thank also the referee, Christopher Johns-Krull,
for his helpful comments which have allowed us to considerably improve
this paper.

\end{acknowledgements}

\end{document}